\documentclass[conference]{IEEEtran}

\usepackage{multicol}
\usepackage[pdftex]{graphicx}
\DeclareGraphicsExtensions{.png}
\usepackage{pifont}
\usepackage{amssymb}
\usepackage{url}

\begin{document}
%
\title{Utility of a Behavlets approach to a Decision theoretic predictive player model}

\author{\IEEEauthorblockN{Benjamin Ultan Cowley}
\IEEEauthorblockA{BrainWork Research Centre,
Finnish Institute of Occupational Health\\
POBox 40, Helsinki 00250, Finland\\
\\
Cognitive Brain Research Group, University of Helsinki, Finland\\
Email: ben.cowley@helsinki.fi}
\and
\IEEEauthorblockN{Darryl Charles}
\IEEEauthorblockA{School of Computing \& \\Information Engineering,\\
University of Ulster,\\
Northern Ireland\\
Email: dk.charles@ulster.ac.uk}
}

\maketitle

\begin{abstract}
We present the second in a series of three academic essays which deal with the question of how to build a generalized player model. We begin with a proposition: a general model of players requires parameters for the subjective experience of play, including at least three areas: \textsf{a}) player psychology, \textsf{b}) game structure, and \textsf{c}) actions of play. Based on this proposition, we pose three linked research questions, which make incomplete progress toward a generalized player model:
\textsf{RQ1} \textit{what is a necessary and sufficient foundation to a general player model?};
\textsf{RQ2} \textit{can such a foundation improve performance of a computational intelligence-based player model?}; and
\textsf{RQ3} \textit{can such a player model improve efficacy of adaptive artificial intelligence in games?}
We set out the arguments for each research question in each of the three essays, presented as three preprints.
The second essay, in this preprint, illustrates how our 'Behavlets' method can improve the performance and accuracy of a predictive player model in the well-known Pac-Man game, by providing a simple foundation for areas \textsf{a}) to \textsf{c}) above.
We then propose a plan for future work to address \textsf{RQ2} by conclusively testing the Behavlets approach. This plan builds on the work proposed in the first preprint essay to address \textsf{RQ1}, and in turn provides support for work on \textsf{RQ3}.
The Behavlets approach was described previously; therefore if citing this work please use the correct citation:

Cowley, B., \& Charles, D. (2016). Behavlets: a Method for Practical Player Modelling using Psychology-Based Player Traits and Domain Specific Features. \textit{User Modeling and User-Adapted Interaction}, 26(2), 257-306.
\end{abstract}


\section{Introduction}
\label{intro}
We argue that for generalised game AI to play at a human-level will require a model of player psychology.
Such a generalised player model requires parameters to describe facets of the player's subjective experience, drawn from a foundation of established models, including at least: \textsf{a}) psychology of behaviour; \textsf{b}) general game design; and \textsf{c}) actions in the context of a given game. This foundation should be integrated with the computational intelligence that drives the model.

These arguments imply several research questions. In the first preprint in this series, \cite{Cowley2016pre1}, we discussed how to improve the theoretical validity of such a foundation by meta-analysis. In this second preprint in the series, we discuss \textsf{RQ2}: \textit{can such a foundation improve algorithmic performance of the computational intelligence required for a real-time player model?}

In this preprint we give a proof-of-principle that such a foundation, based on 'Behavlets' approach for linking psychology to game-play, can practically improve a simple predictive model of Pac-Man players. We previously proposed the Behavlets method to build facets \textsf{a}) to \textsf{c}) above into composite features of game-play defined over entire action sequences \cite{Cowley2016behavlet}, and thus model players for, e.g. personality type classification \cite{Cowley2013}. The Behavlet approach is briefly recapped below.

Subsequently in Methods section we report a controlled comparison of the use versus non-use of Behavlets, on two Decision Theory models for predicting player movement in Pac-Man (implemented in C++ Direct X). One model uses simple features calculated for a single state of the game, and the other uses Behavlet-like composite features. Decision theory fundamentals are described below. As described in Results section, the latter model improves speed (from non-real-time to real-time), and accuracy by 35\%.

However to \textit{comprehensively} address \textsf{RQ2} will require further work, by adding a validated foundation to a general player model, both topics for future work as described in Discussion. An empirically-supported answer to \textsf{RQ2} will advance work on the third planned question, \textsf{RQ3} \textit{can such a player model improve efficacy and viability of the AI required to power games which adapt to their players?} (addressed in the third preprint in this series \cite{Cowley2016prep3}).

\subsection{Behavlets background}
\label{behav}
For full details of the Behavlet process see \cite{Cowley2016behavlet}; from that paper, the following process description outlines an iterative process for a game designer/developer (GD):

\begin{enumerate}
	\item Gameplay Analysis and Mapping:
	\begin{enumerate}
		\item game structure and context (\textit{leads to}) \ding{221}
		\item game mechanics and dynamics \ding{221}
		\item game design patterns
	\end{enumerate}
	In this stage, GD should identify which game components give the player \textit{agency}, and are thus central to describing player behaviour. Further, GD must differentiate between game mechanics and the dynamical operations of player-game interaction. Then game-specific design patterns can be identified.

	\item Feature/Behavlet Identification:
	\begin{enumerate}
		\item traits of play behaviour \ding{221}
		\item traits vs patterns \ding{221}
		\item observation and Behavlets
	\end{enumerate}
	In this stage, GD specifies how behaviour would express itself. GD uses a list of descriptive terms for behaviour, to characterise how design patterns would turn into extended sequences of action selection. GD then observes the play of the game to develop Behavlet concepts.

	\item Behavlet coding:
	For the informal Behavlets from step 2, GD then defines pseudo-code and game engine encodings.

	\item Feature selection and testing.
\end{enumerate}

The Behavlets used below are from the same set derived and reported in \cite{Cowley2016behavlet}; readers should thus refer to that paper (esp. appendices) when, e.g., examining Table~\ref{table:loo} below.

\subsection{Decision Theoretic Player Modelling Background}
\label{dtpm}
Predictive player modelling works by considering the player's in-game goals as equivalent to some target function of the game state, parameterized by predefined \textit{utilities}; this function is calculated using observed player data \cite{Thue2006}. In many classes of games, mechanics of play involve choosing the action which maximises a utility function, from a set of actions situated in a possibility-space evolving toward minimal utility in the absence of player action - in other words, act or lose.

Decision Theory \cite{Curley1989} is a formulation of the uncertainty of outcomes due to making non-trivial choices, which adapts well to modelling game-play. Working from the Decision Theory formulation for game-play by \cite{Gmytrasiewicz2000}, we define our own as follows.

A rational player makes decisions by picking from a finite set $A$ of alternative courses of action. $a \in A$ can be thought of as a plan consisting of consecutive moves extending to the future time $t_a$. The limit on this \textit{look-ahead time} will be $t_{max}$\footnote{In a game of Pac-Man $t_a$ will almost always equate $t_{max}$ since for any one plan, only level-end or Pac-Man's death results in cessation of planning}. An action plan takes place in the set $S$ of all possible game states. So each $a$ corresponds to a sequence of states $s \in S$ starting with a state 'adjacent' to the current state and ending with $s^{t_a}$. Since all the states considered in each decision-making situation are limited by $t_{max}$ they form a subset of $S$ which we call $S_t$. To obtain the necessary ordering of $s$ when selecting from $S_t$ so that the sequence $a$ makes sense (since $S_t$ is unordered), we identify each state by its distance in the future, i.e. $s^t$.

We represent player uncertainty with a time-wise probability function giving a distribution $P(S)$ over $S$. This function is a temporal projection expressed as $proj:S \times A \to P(S)$; such that the action $a$ given the current state $s^0$ results in the probability of the projected states $proj(s^0,a)=P_a (S)$. Specifically, $p_a^t$ will be the probability assigned by $P_a (S)$ to the state $s^t$. The utility function $util:S \to \mathbb{R}$ encodes how desirable the player finds the projected states.

Using this general notation for Decision Theory in some method-specific formulation of probabilities and utilities, we can predict the maximum utility plan $a$ that a player should perform. The two player modelling approaches compared below each encode their characteristics as a specific formula.

\section{Methods}
\label{methods}
To test the two models, we collected a data set of games for testing, using the same methodology as described in \cite{Cowley2016behavlet}, including ethical approval from the Research Governance board of the University of Ulster. 37 players participated; each played a number of practice games which were excluded; and generated 105 test games from two to three post-practice plays each. Two tree-search modelling approaches were compared in this experiment: a '\textit{simple-features}' model and a '\textit{Behavlet-based}' model. Our hypotheses for the comparison were that the Behavlet model would outperform the simple model in terms of: \textsf{H1} \textit{accuracy}, because the Behavlets better capture the player's intentions; \textsf{H2} \textit{speed}, due to the reduced number of computations required (as explained below); and \textsf{H3} \textit{insights}, because an explanatory framework is built into the Behavlet features.

\subsection{Tree Search Preliminaries}
In this tree-search version of Decision Theory, the meaning of some symbols is refined. Thus the subset of states $S_t$ corresponds to the finite look-ahead tree of future states described below, and $t_{max}$ is the computational limit on tree size. A plan $a$ corresponds to a path in the tree, with the last state $s^{t_a}$ equating a leaf which can uniquely identify the path.

A 'classic' look-ahead tree is built by calculating all possible combinations of positions that the in-game actors can occupy in one 'step', and then iterating for a computationally tractable number of steps. Building the tree explores the game's \textit{possibility space}, ranking each potential future state by calculating the utility to the player of the features found in that state. To calculate the utility weights for the tree, the ideal metric in any game would be the difference between the value of some utility for the current state, and the value for this utility in the final state \cite{Samuel1959}. Since each possible state at a time-step is ranked by its utility contribution to the path to which its parent belongs, the algorithm navigates the tree of states along the path of highest utility in order to 'back-up' the prediction of which next move is optimal.

The tree \textit{branching rate} corresponds to the number of possible moves available to each actor at each step. Thus branching rate is relevant to computational tractability. Our Pac-Man map has 143 navigable squares with two adjacent squares; 32 with adjacency three; and seven with adjacency four. Therefore, if we consider the future moves of a number of actors $w$, the minimum rate would be given by equation~\ref{eq.branch}:

\begin{equation}
	\frac{143}{182}.2w + \frac{32}{182}.3w + \frac{7}{182}.4w \approx 2.25w
	\label{eq.branch}
\end{equation}

In practice the rate is higher because squares with adjacency three or four tend to be entered more often. In our test data we estimate the actual branching rate is $\sim$2.75$w$ on average. Given this rate we can calculate a default accuracy of 36\% generated by random choice of next move.

In both models, two parameters are important to help conserve computation: depth of the look-ahead tree; and a heuristic of player behaviour which we term the 'maximum back-tracking limit'. This heuristic places a limit $\lambda$ on the number of moves a player can make along a bi-directional corridor before the algorithm ceases to consider the backwards direction in its predictions. Thus, leaves of the tree will be pruned if they extend to the direction the player has come from more than $\lambda$ moves ago. The premise behind this heuristic is that players are goal-directed.

We conducted a parameter sweep of depth and $\lambda$ for each model. The range of each parameter was bounded, depth from 4..9 and $\lambda$ from 3..6. The lower-bound on depth arose because for some Behavlets, estimation from three or less states would be ill-defined. The upper-bound on depth was fixed to limit computation time: $\sim$14 seconds per state was required for depth nine tree-search, totalling $\sim$32 hours per test game. The lower bound on $\lambda$ was set to allow for backtracking by human error; the upper bound on $\lambda$ was set to the maximum length of game map corridors. The parameter sweep was conducted in a similar manner for each model, over the same data set. Both models had best accuracy at depth=4 and $\lambda$=5.

\subsection{Model 1}
\begin{figure*}[!ht]
	\centering
	\includegraphics[width = 5in]{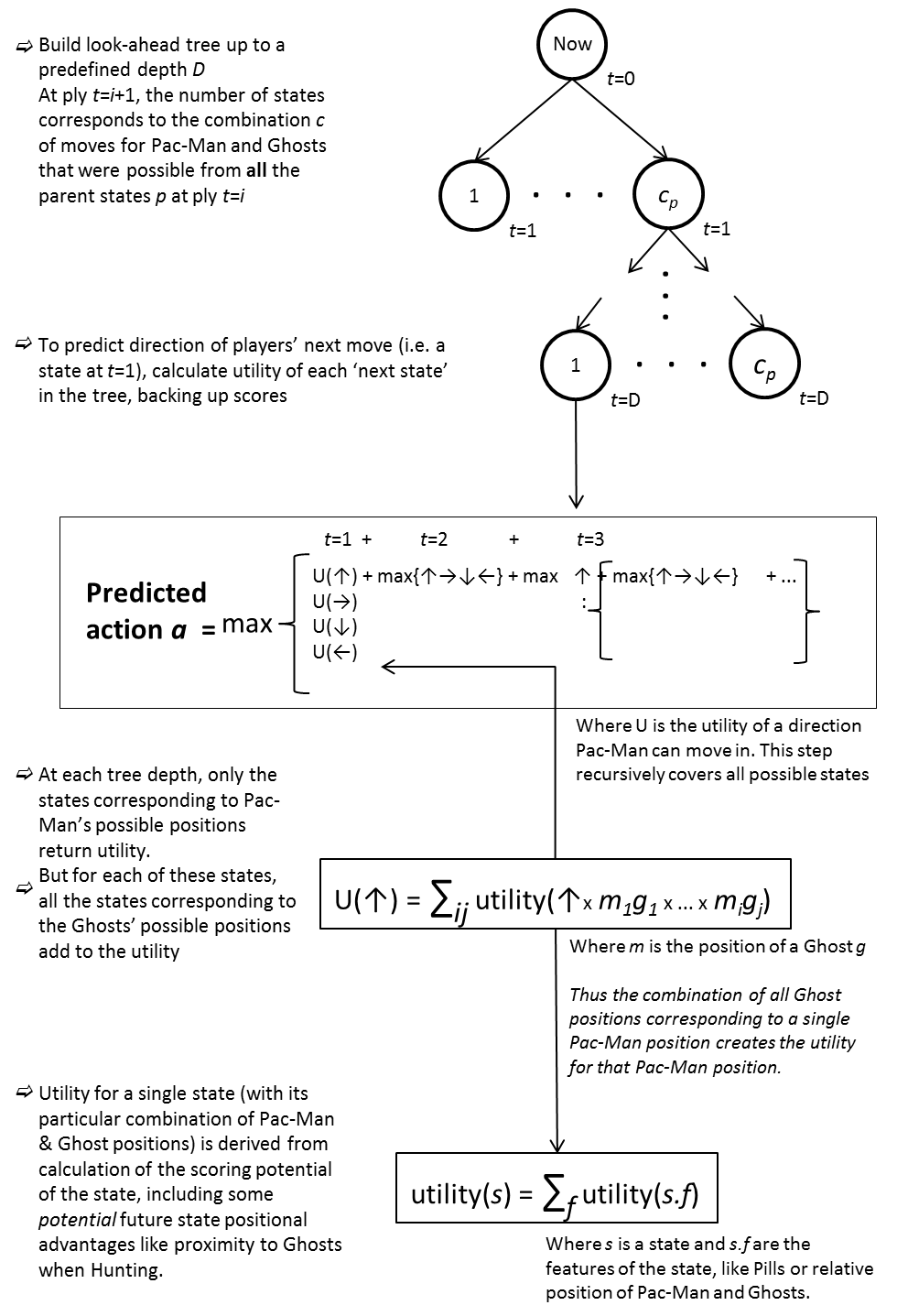}
	\caption{Schematic of the operation of the model 1 algorithm. All possible future moves are calculated, by generating a look-ahead tree of depth D. The utility of each state in each path is then accumulated to give a final score to each possible direction of movement, allowing a prediction for action $a$}
	\label{fig.model1}
\end{figure*}

Model 1, termed the \textit{simple-features} approach, performs classic tree-search, calculating cumulative utility of simple heuristic features in each state in a path. Since the look-ahead tree was extensive, it was run off-line. The operation of model 1 is described schematically in Figure~\ref{fig.model1}, and defined in formula~\ref{eq.model1}:

\begin{equation}
	a_t^* = ArgMax(a \in A) \sum_{s \in S_t} p_a^t \times util(s)
	\label{eq.model1}
\end{equation}

The heuristic features are based on the items and events which reward or threaten the player, i.e. are of importance in the game. These include Pac-Man and Ghosts; \textit{Pills} that the player must collect to pass a Level; Power Pills that switch the roles of Ghosts from hunters to hunted; and Fruit that acts as a bonus reward.

\begin{itemize}
	\item Threat - Ghost proximity (measured by A* algorithm) and distribution across the map.

	\item Reward - count of each Pill weighted by the number of adjacent Pill and the inverse of distance to Pac-Man.

	\item Number of Lives left to Pac-Man.

	\item 'Hunt' reward (when Pac-Man has eaten a Power Pill) –
	\begin{itemize}
		\item If game is in Hunt mode, this is just Ghost proximity.

		\item If not, then this is proximity to nearest Power Pill combined with Ghost proximity.
	\end{itemize}
\end{itemize}

\subsection{Model 2}
In model 2 Behavlets contribute to utility calculation for each \textit{path} from the look-ahead tree, in contrast to model 1 which sums the utilities of every \textit{state} in the tree. For this model, Behavlets were adapted to work over sequences of states where some proportion of states are \textit{predicted}. Adapted Behavlets retained the core logic defined in \cite{Cowley2016behavlet}, but we excluded i) any Behavlet defined only over a long sequence, e.g. game level; ii) Behavlets with logic incompatible with prediction, e.g. those based on speed of movement. 

Focusing the model on the player's perspective using Behavlets allows an simple yet effective optimization: to the player, future Ghost positions are estimated as a probability distribution. Branching only for the potential moves of Pac-Man, and not the Ghosts, the branching factor is reduced by \textgreater2.25 for every Ghost. Behavlets are calculated using Ghost locations estimated from their movement probability distribution; model 2 can thus discard the $proj$ function and avoid calculating exhaustive look-ahead trees.

Model 2 also dynamically adjusts Behavlet use with a 'state-checker': each tree search is constrained to the Behavlets contextually relevant to the game state.

Formula~\ref{eq.model2} for model 2 retains the established definitions, but calculates utility for an entire sequence-of-states or plan $a$, rather than just one state. Plan $a$ is selected from $S_t \subset S$, defined by the current state and the computational limit $t_{max}$. $a$ defines all \textit{included} Behavlets $F_a$, where $f \in F$ the set of all Behavlets, as input to $util$. Thus $util:S \times A \to \mathbb{R}$ assigns value to states; summed output of multiple $f \in F_a$ gives the utility score of a plan $a$: best scoring $a$ predicts next move.

In look-ahead tree terms, model 2 calculates utility for an entire look-ahead tree path from Behavlets which trigger for that path. The model 2 algorithm is described schematically in Figure 3.

\begin{equation}
	a_t^* = ArgMax(a \in A)  util(\sum_{f \in F_a} f)
	\label{eq.model2}
\end{equation}

\begin{figure*}[!ht]
	\centering
	\includegraphics[width = 5in]{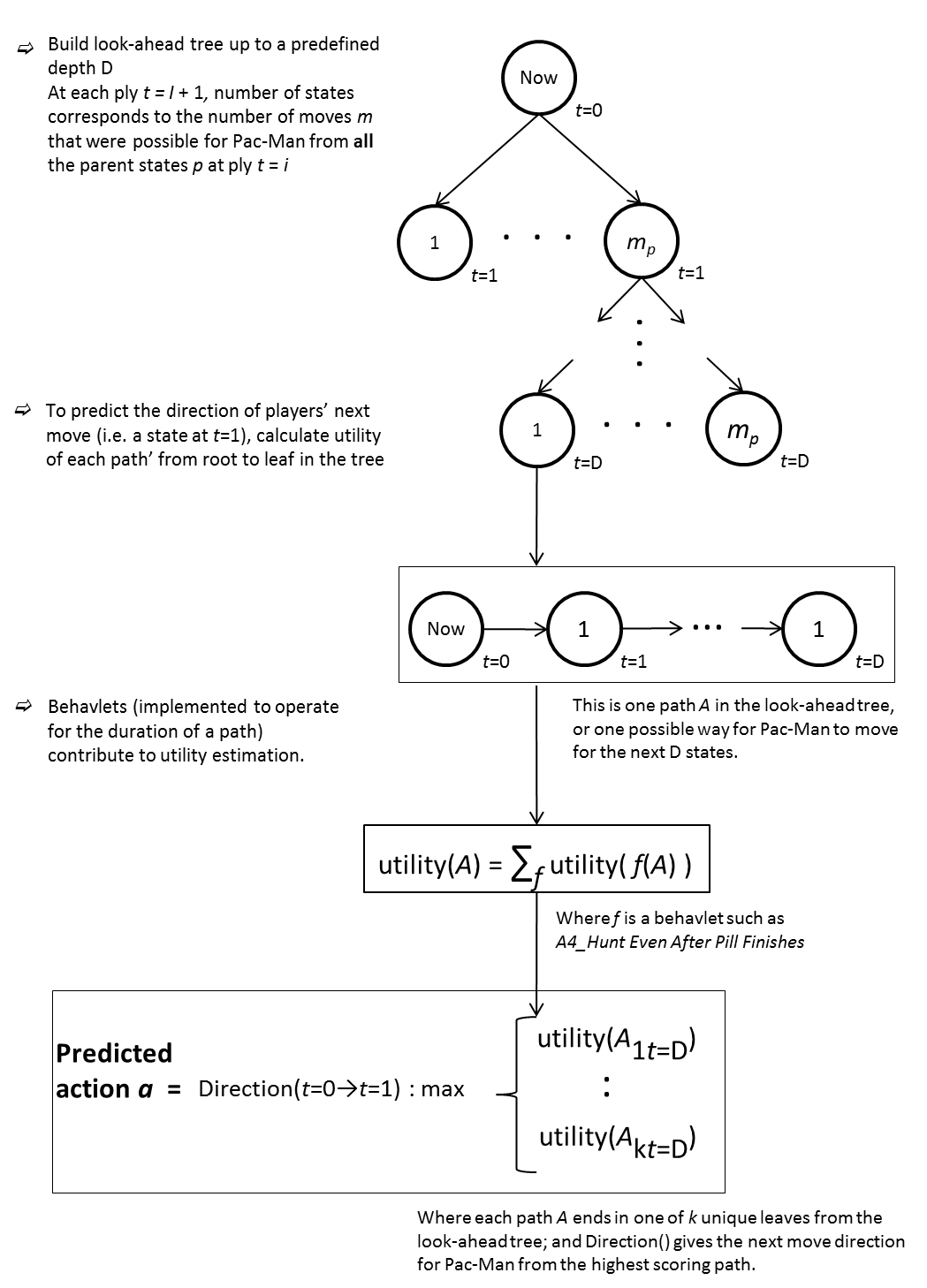}
	\caption{Schematic of the algorithm for model 2. Multiple Behavlets contribute to utility calculation for each path in the tree, to evaluate action $a$}
	\label{fig.model2}
\end{figure*}

\section{Results}
\label{results}
Model 1 had a prediction accuracy average of 39\%, 3\% above the random chance default accuracy of 36\%. This meant that for games which had an average length of 327 states, on average only 131 moves were correctly predicted. The number of consecutive predictions (a sign of accurate classification of activity sequences) averaged 2.3 moves in length, with standard deviation of 2.4, implying that model 1 does not predict long sequences of Pac-Man's actions.

Model 2 accuracy was 70.5\%: this represents a lift of $\sim$35\% from the random chance default of $\sim$36\%. The millisecond speed of execution per state was M=81, md=63, SD=37. Given that the state-rendering rate of our Pac-Man engine was 10-11 Hz at 96ms per frame, this performance allowed real-time execution.

Comparing the models, the accuracy difference between 39\% and 71\% supports \textsf{H1}. To clarify whether Behavlets or the difference in algorithm is responsible for improved accuracy, we performed a simple 'leave-one-out' test, summarised in Table~\ref{table:loo}. The test operates by measuring a baseline for speed and accuracy including all Behavlets and state-checker. Then each Behavlet is iteratively excluded from the model. Compared to baseline, if exclusion of Behavlet $i$ raises execution time (indicated by a negative difference) and lowers accuracy (positive difference), then Behavlet $i$ should improve speed and accuracy. The second row of the table tests the removal of state-checker code. Cutting it means using all Behavlets in any given prediction, increasing execution time to 296ms.

The columns of Table 1 are: name of the \textbf{Excluded} Behavlet, or state-checker; \textbf{ms/State} millisecond computation time per state \textit{difference from baseline}; \textbf{Acc\%} accuracy \textit{difference from baseline}; and \textbf{Usage}, if the Behavlet should be included in a final model, Yes or No.

\begin{table}[ht]
	\centering
	\caption{Comparative speed and accuracy results of excluding each feature in turn.}
	\label{table:loo}
	\begin{tabular}{lccc}
		\textbf{Excluded} & \textbf{ms/State} & \textbf{Acc \%} & \textbf{Usage} \\
		None – baseline                         & 240 & 72.7 & -\\
		State-checking code$^1$                 & -56 & 0.5 & Y \\
		Points\_Max                             & -4 & 16.2 & Y \\
		A1\_Hunt Close To Ghost House           & -4 & 0 & Y \\
		A4\_Hunt Even After Power Pill Finishes & -31 & 0 & Y \\
		A6\_ Chase Ghosts or Collect Dots       & -13 & 0.1 & Y \\
		C1.b\_Times Trapped and Killed          & 42 & 0 & N \\
		C2.a\_Average Distance to Ghosts$^2$    & -1 & 0.3 & Y \\
		C2.b\_Average Distance During Hunt      & -23 & 1.6 & Y \\
		C3\_Close Calls                         & -20 & 0.1 & Y \\
		C4\_Caught After Hunt                   & -40 & -0.1 & Y \\
		C5\_Moves With No Points Scored         & 0 & 0.9 & Y \\
		C7\_Killed at Ghost House               & -16 & 0.1 & Y \\
		Cherry Onscreen Time                    & 2 & -0.3 & N \\
		D2\_Player Vacillating                  & -11 & 0.1 & Y \\
		P1\_Wait Near Power Pill to Lure Ghosts & 120 & 0.1 & N \\
		P1.c\_Lure: \# Ghosts Eaten After Lure  & 120 & 0.1 & N \\
		P1.d\_Lure: Caught Before Eating Pill   & 120 & 0.1 & N \\
		P4\_SpeedOfHunt                         & -5 & 0 & Y \\
		S2a\_Lives Gained                       & 10 & -0.2 & N \\
		S2b\_Lives Lost                         & 10 & -0.2 & N \\
		S4\_Teleport Use                        & 11 & 0 & N
	\end{tabular}
	\flushleft{1 - determines use of state-dependent features\\
		2 - outside of Power Pill mode only}
\end{table}

An important result here is that the \textit{Points\_Max} feature reduces accuracy by $\sim$16\% when excluded. The Behavlet exclusions affected accuracy by 0-2\%. Thus about half the lift over default is due to this single feature. However the other half is attributable to the combined Behavlets, and the fact that no single Behavlet dominates suggests that all are contributing, perhaps by interaction. The importance of \textit{Points\_Max} can be attributed to frequency: \textit{Points\_Max} influences every single utility calculation, while e.g. \textit{C3\_Close Calls} might only fire a few times per level or per game.

Hypothesis \textsf{H2} is well supported by the increase in performance to real-time. Dedicating more resources to utility calculation does not seem to be a requirement for accurate predictions. There is a positive linear relationship between accuracy and longer execution times, shown in Figure~\ref{fig.speedacc} below, but the Pearson correlation coefficient of 0.16 is non-significant, p $\simeq$ 0.1.

\begin{figure}[!ht]
	\centering
	\includegraphics[width = 2.5in]{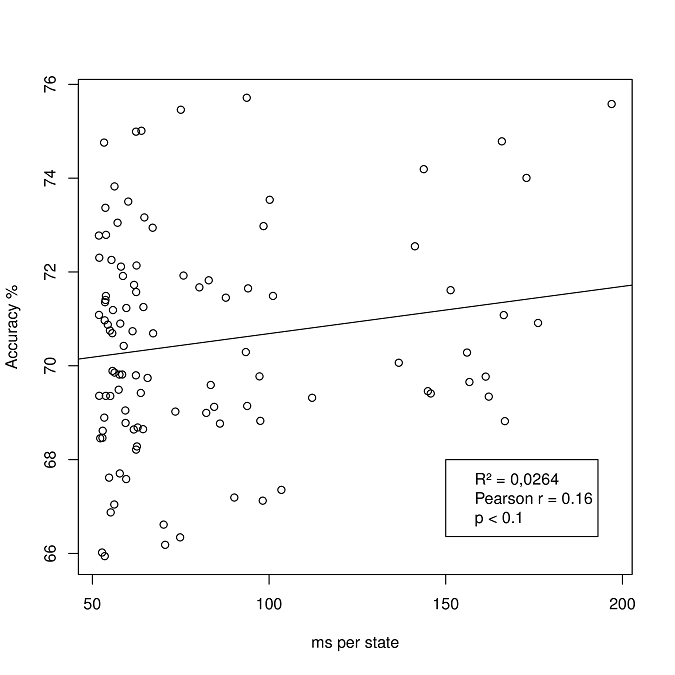}
	\caption{Scatter plot of individual test games, with linear relationship between accuracy and speed of execution.}
	\label{fig.speedacc}
\end{figure}

To address hypothesis \textsf{H3}, we examine the correlation between results from model 2 and our work on player type classification \cite{Cowley2013}, where Behavlets were heavily used to generate insights \footnote{Details of this paper cannot be reproduced as licence is not open access.}. The same 37 players participated in each study, and we compared the classification score they received in the earlier study to their mean 'Behavlet-based' prediction accuracy. The two sets of results share a Pearson correlation coefficient of 0.5, \textit{p}=0.001 (two-tailed). The type classifier was a continuous scale between two class labels, termed \textit{Conqueror}/\textit{Not Conqueror}; thus, the correlation indicates that model 2 had better accuracy for Conqueror type. This suggests that Behavlet player type relationships proven in \cite{Cowley2013} can be used to reason about game periods with accurate predictions.

\section{Discussion}
\label{disc}
Comparison of Behavlets with a 'na{\"i}ve' approach, and earlier work \cite{Cowley2013}, demonstrates some of the value of Behavlets for player modelling. We believe that more can be gained from the Behavlet approach by fixing some existing limitations. For example, model 2 implies that the algorithm can be biased toward any given pre-classified type, by tuning utility-weights for each Behavlet with respect to correlated type scores. In fact, a simple hill-climbing algorithm for weight tuning was reported in \cite{Cowley2009a}, illustrating that simple solutions could address this issue. For more Behavlet issues see \cite{Cowley2016behavlet}.

To advance the general player modelling approach, future work will study the efficacy of building state-of-art machine learning models on a foundation of psychology, game design patterns and player action preferences. This foundation will be an evolution of Behavlets informed by the theoretical validation work defined in the previous preprint, \cite{Cowley2016pre1}. Implementation will include both a modern, commercial-standard game for ecological validity, and a formally well-defined game to facilitate rigorous analysis under a formal model. These implementations will help address the third planned question, \textsf{RQ3}, to investigate whether such general models can not only help understand players, but drive AI to play responsively with players.

\section{Conclusion}
\label{conc}
We presented a comparative study of two predictive player models in Pac-Man. The outcome shows conclusive improvement with the Behavlet foundation, and suggests the potential for further studies on adding such psychological foundations to computational intelligence for player models.

\section*{Acknowledgements}
Partly supported by Tekes, the Finnish Funding Agency for Innovation, project Re:Know \#5159/31/2014


\bibliographystyle{abbrv}
\bibliography{Cowley_Behavlet_preprint_2_bib}


\end{document}